\documentclass[useAMS,usegraphicx,usenatbib]{mn2e}
\bibpunct{(}{)}{;}{a}{}{,}
\newcommand{\msun}{{\rm M}_{\odot}}

\newcommand{\xte}{{\it RXTE}}
\newcommand{\pca}{{\it RXTE}/PCA}
\newcommand{\asm}{{\it RXTE}/ASM}

\newcommand{\cx}{\mbox{Cyg~X-3}}
\newcommand{\exo}{\it EXOSAT}

\begin{document}
\title[X-ray variability of Cygnus X-3]{The aperiodic broad-band X-ray variability of Cygnus X-3}
\author[M.~Axelsson, S.~Larsson, and L.~Hjalmarsdotter]{Magnus Axelsson,\thanks{E-mail: magnusa@astro.su.se} Stefan Larsson, and Linnea Hjalmarsdotter\\ 
Department of Astronomy, Stockholm University, SE- 106 91 Stockholm, Sweden}

\maketitle

\begin{abstract}
We study the soft X-ray variability of Cygnus X-3. By combining data from the All-Sky Monitor and Proportional Counter Array instruments on the {\it RXTE} satellite with {\it EXOSAT} Medium Energy (ME) detector observations, we are able to analyse the power density spectrum (PDS) of the source from $10^{-9}-0.1$ Hz, thus covering time-scales from seconds to years. As the data on the longer time-scales are unevenly sampled, we combine traditional power spectral techniques with simulations to analyse the variability in this range.
The PDS at higher frequencies ($\ga 10^{-3}$ Hz) are for the first time compared for all states of this source. We find that it is for all states well-described by a power-law, with index $\sim -2$ in the soft states and a tendency for a less steep power-law in the hard state. At longer time-scales, we study the effect of the state transitions on the PDS, and find that the variability below $\sim 10^{-7}$ Hz is dominated by the transitions. Furthermore, we find no correlation between the length of a high/soft state episode and the time since the previous high/soft state.
On intermediate time-scales we find evidence for a break in the PDS at time-scales of the order of the orbital period. This may be interpreted as evidence for the existence of a tidal resonance in the accretion disc around the compact object, and constraining the mass ratio to $M_2/M_1\la0.3$. 
\end{abstract}


\section{Introduction}

The persistent X-ray source Cygnus X-3 was discovered 40 years ago and is one of brightest and most extensively studied X-ray binary (XRB) systems. It is however a peculiar source that still eludes simple classification. It is a close binary with an orbital period of 4.8 hours \citep{par72}. The identification of the donor as a Wolf-Rayet star \citep{vke92} classifies it as a high-mass X-ray binary (HMXB). The strong wind from the Wolf-Rayet companion, surrounding the entire system makes radial velocity measurements difficult and the masses of the components are therefore uncertain. Recent studies based on the luminosity at the observed state transitions suggest that the accretor is a $\sim30\,\msun$ black hole \citep{hja08a,hja08b}. This would make {\cx} similar to the two other Wolf-Rayet X-ray binaries discovered so far, IC 10 X-1 \citep{pre07,sil08} and NGC 300 X-1 \citep{car07a,car07b}, where at least one of them, IC 10 X-1, contains a black hole of mass 23--34$\msun$.

\subsection{Spectral states}

{\cx} displays five distinct spectral states \citep{sz04}, similar to those of the XRB transients ({\cx} itself is however a persistent source), but with the hardest state peaking at much lower energies. The presence of strong absorption has made it difficult to interpret the intrinsic spectral shapes and whether they represent true state transitions or are merely an effect of variable local absorption. \citet{hja08a}, however, presented several strong arguments that the transitions indeed correspond to an intrinsic change in the accretion geometry, and in \citet{hja08b} {\xte} data from all spectral states were consistently modelled. In this paper we use the nomenclature of \citet{hja08b} to describe the states as the hard, intermediate, very high, soft non-thermal and ultrasoft (see figs.~1 and 7  in \citealt{hja08b} for spectral shapes and envisaged accretion geometries) in order of increasing softness. The very high, soft non-thermal and ultrasoft states will for most of the time be combined and referred to as the ``soft state'' or states as they differ mainly in the shape and strength of the high-energy tail.

\subsection{Variability}

The X-ray temporal properties of {\cx} are not as well studied as those of for example Cyg~X-1. 
The strong orbital modulation with a period of 4.8h was discovered in 1972 \citep{par72}. In a study of the short time variability using {\exo} data, \citet{vdk85} reported transient quasi-periodic features in the lightcurves during the soft states, and possibly also in the hard state. The periods for these features were in the 50--1500\,s range, with a given feature surviving for 5--40 periods. Similar results were reported by \citet{rao91}, who found a transient oscillation with a $\sim121$\,s period.

A power density spectrum (PDS) of {\cx} was presented by \citet{wil85}, covering the frequency range $\sim10^{-5}$--\,$0.1$\,Hz. It was found that it was well described by a power-law of index $-1.8$, with a flattening at lower frequencies. A strong component was seen from the orbital modulation. No variability signal was detected above 0.1\,Hz. A more detailed study of the X-ray variability above 1\,Hz by \citet{ber94} found an upper limit of 12\% rms on the variability in this range.

Recently, \citet{cho04} presented results from analysing {\pca} observations of {\cx}, including corrections for the binary variations. They compare observations made at three different times, and find that the PDS does not change significantly; above $10^{-4}$\,Hz it can be fitted by a power-law with index $-1.5$. Using {\asm} data, they find that the power-spectrum on longer time-scales ($>1$ day) is flatter, and through extrapolation predict a break around $10^{-4}$\,Hz.

In this paper we study the PDS of {\cx} in the range $10^{-9}-0.1$ Hz using {\pca}, {\asm} and {\exo} data. Based on the spectral analysis of \citet{hja08b} we look for systematic differences in the PDS between the different states. We also study the long-term variability and the full broad-band PDS between the shortest and longest time-scales available. The data and methods used are described in Sect.~2, and the results are presented in Sect.~3. This is followed by a discussion in Sect.~4, and in Sect.~5 we summarise our conclusions.

\section{Data and analysis}
The bulk of our data is from the All-Sky Monitor \citep[ASM,][]{lev96} and Proportional Counter Array \citep[PCA,][]{jah96} instruments onboard the {\it RXTE} satellite. The {\asm} provides continual coverage of the X-ray sky in the 2--12\,keV range. A given source will be observed several times each
day, and the data is available either in dwell-by-dwell format, or one-day averages. The {\pca} is used
for pointed observations, typically lasting up to about an hour. The pointing data in this paper is the same as in \citet{hja08b}, and that spectral analysis is used when determining the state of the source.

To create the PDS, PCA lightcurves were extracted with 100\,ms resolution in the 2--9\,keV band. The lightcurves were then separated in intervals of 4096 bins, and a PDS was computed for each interval. Only intervals without gaps were retained, and the resulting PDS were then averaged to form the final power spectrum. Since the data stem from many different observations with varying setup, the Proportional Counter Units (PCUs) in use vary between observations. All lightcurves have therefore been normalized to one PCU.

To bridge the gap in frequency coverage between the PCA and ASM datasets, we have also included data from the {\exo}/Medium Energy (ME) instrument \citep{tur81}. The data comprise the 19 observations of {\cx} made with {\exo} spanning more than 9\,ks (in total 22 observations were made of {\cx} between 1983 and 1985, see \citealt{ber94} for a log). Due to gaps in the data, these observations were divided into 26 segments, spanning between 3000 and 27\,750\,s. The data cover the energy range 0.8--8.9 keV. Included in this dataset is the long (106\,ks) observation analysed by \citet{wil85}.

X-ray lightcurves of {\cx} show strong modulation at the 4.8 hour orbital period. We have corrected all
PCA, ASM dwell-by-dwell and {\it EXOSAT} lightcurves for this modulation by dividing fluxes with a modulation template, taken from \citet{vdk89}. The amplitude of the template was fitted to the data
with phases given by the ephemeris in \citet{vdk89}.

A problem with interpreting ASM data is dealing with systematic effects. The data show variations with periodic components on time-scales matching one year, one day and one orbit. These periods are related both to variations in the observing conditions and in the instrument properties. Although correction factors are applied when producing the ASM lightcurve data, it is not possible to completely eliminate these effects. As a result, some systematic effects will still remain; however, for a bright source such as {\cx}, these systematics are only present on a low level for long time-scales ($\la 3$\,\%; A. Levine 2008, priv. comm.). 

\subsection{Unevenly sampled data}
\label{uneven}

Although the observations made by the {\asm} are combined into one-day averages, they are not evenly sampled. A given source will be observed several times per day, in principle allowing variability on shorter time-scales to be detected by analysing the dwell-by-dwell data. In the case of Cyg X-3 it also allows us to correct the data for the orbital modulation.

The {\asm} data was split up into different states, based on the flux level. Small pieces of data in-between the defined states were not used. The data for each state was then divided into segments with
lengths between 250 and 1000 points. Power density spectra were computed with a Lomb-Scargle algorithm, normalized to variance per intensity squared and frequency unit. A logarithmically
averaged PDS was computed and a power-law fitted to the frequency range $2\times10^{-7}$--\,$6\times10^{-6}$\,Hz.

The systematic effects on the PDS caused by the data sampling were investigated by Monte Carlo simulations. Synthetic data sets, with the same time sampling as the observations were produced and analysed in the same way. The simulated lightcurves were produced by shot noise. By injecting exponential shots with different distributions in decay times we simulated lightcurves for which
the PDS were power-laws with index ranging from $-1$ to $-2$. The difference between the simulated and measured power-law index allowed us to estimate the systematic errors introduced by the data sampling.

\section{Results}

\subsection{PDS at high frequencies: ${\bf 10^{-3}-0.1}$\,Hz}

\begin{figure}
\includegraphics[width=8cm]{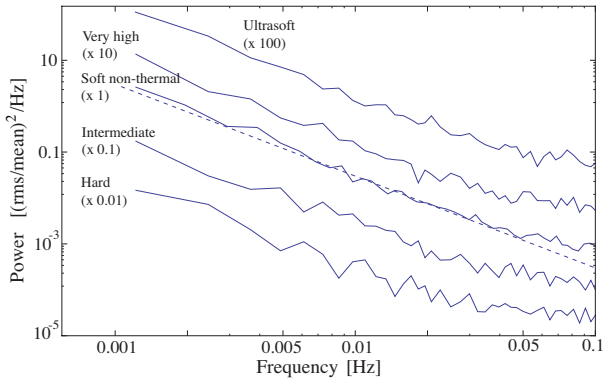}
\caption{Power spectra from {\pca} data for all states of Cyg~X-3, shifted for clarity. The dashed line is a power-law with index $-2$. The flattening at high frequencies is due to the Poisson level not having been subtracted.}
\label{pcapds}
\end{figure}

Using {\pca} data, we have conducted a systematic study of the high frequency PDS of {\cx}. Figure~\ref{pcapds} shows example PDS for all states seen in the source as defined by \citet{hja08b}. The PDS is in all cases well described by a steep power-law up to $\sim 0.1$ Hz. At higher frequencies, the signal is too weak to be detected. 

\begin{figure}
\includegraphics[width=8cm]{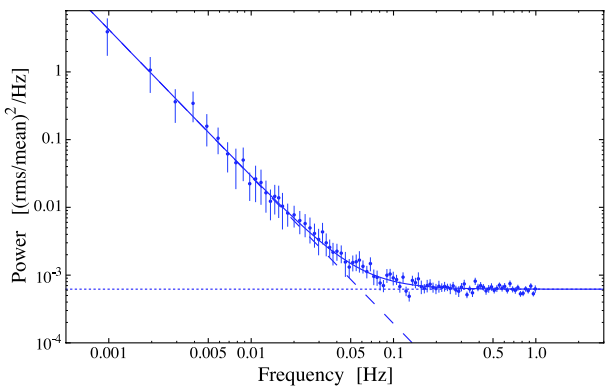}
\caption{Example fit of the PDS of Cyg~X-3 in the soft non-thermal state (solid line). Rather than subtracting the  Poisson level, we used a model comprising a constant component (dotted line) and a power-law (dashed line) to fit the data.}
\label{pcafit}
\end{figure}

In order to test for systematic differences, we fit each PDS to determine the power-law index. Rather than subtract the Poisson level, we include it in the model as a constant. An example fit is shown in Fig.~\ref{pcafit}. 

The resulting indices are then grouped according to state. This allows us to compare the variations within each state to that between states. Table \ref{indextable} lists results and uncertainties for each state, obtained by combining all individual fits for that state. As the number of observations in each state varies, the rightmost column gives the number of PDS combined in each state. We see that the index in all states is close to $-2$. However, there appears to be a tendency for the power-law in the hard state PDS to be slightly less steep. In this state we also noted that there were larger fluctuations between the individual observations. The results are consistent with the index being the same in all the soft states. 

\begin{table}
\begin{center}
\begin{tabular}{l l l c}
\hline \hline
State & Index & $\sigma$ & No. of PDS \\
\hline
Hard & -1.865 & 0.056 & 8 \\
Intermediate & -2.067 & 0.040 & 5 \\
Soft non-thermal & -2.170 & 0.105 & 1 \\
Very High & -2.142 & 0.042 & 8 \\
Ultrasoft & -2.118 & 0.029 & 11 \\

\hline
\end{tabular}
\caption{Fitted power-law indices in the PDS of all states. The data are from {\pca}, and the frequency range is $10^{-3}$--0.1 Hz.}
\label{indextable}
\end{center}
\end{table}

A power-law with index $-2$, sometimes referred to as red noise, can result from a random walk process. Our results could thus indicate that such a process dominates the high-frequency power spectrum in all states of the source, drowning any intrinsic variability. This will be further addressed in Sect.~\ref{wind}.

\subsection{Long time-scales: ${\bf 10^{-9}-10^{-5}}$\,Hz}

\begin{figure*}
\includegraphics[width=16cm]{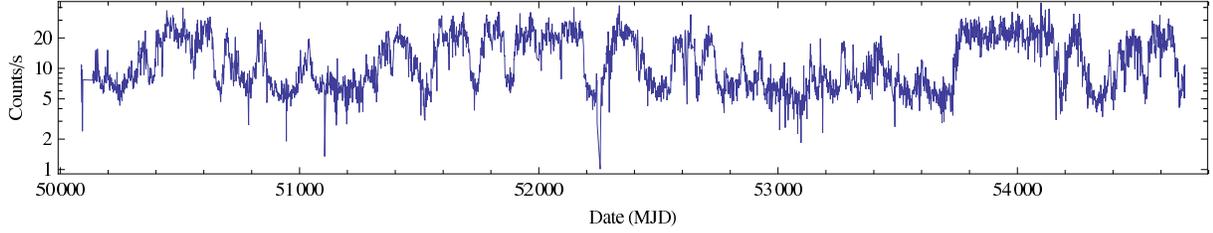}
\caption{2--12\, keV lightcurve of Cyg~X-3 . The data span more than 12 years, from March 1996 to July 2008. The soft states are clearly visible as periods of increased flux.}
\label{asmlc}
\end{figure*}

The {\asm} has observed {\cx} for over 12 years, giving unprecedented opportunity to study the
long-term behaviour of the source. Figure~\ref{asmlc} shows the 2--12\,keV lightcurve of {\cx}. 
The state transitions between the hard state with low ASM flux and the soft states with high ASM flux are clearly visible. 

\begin{figure*}
\includegraphics[width=8cm]{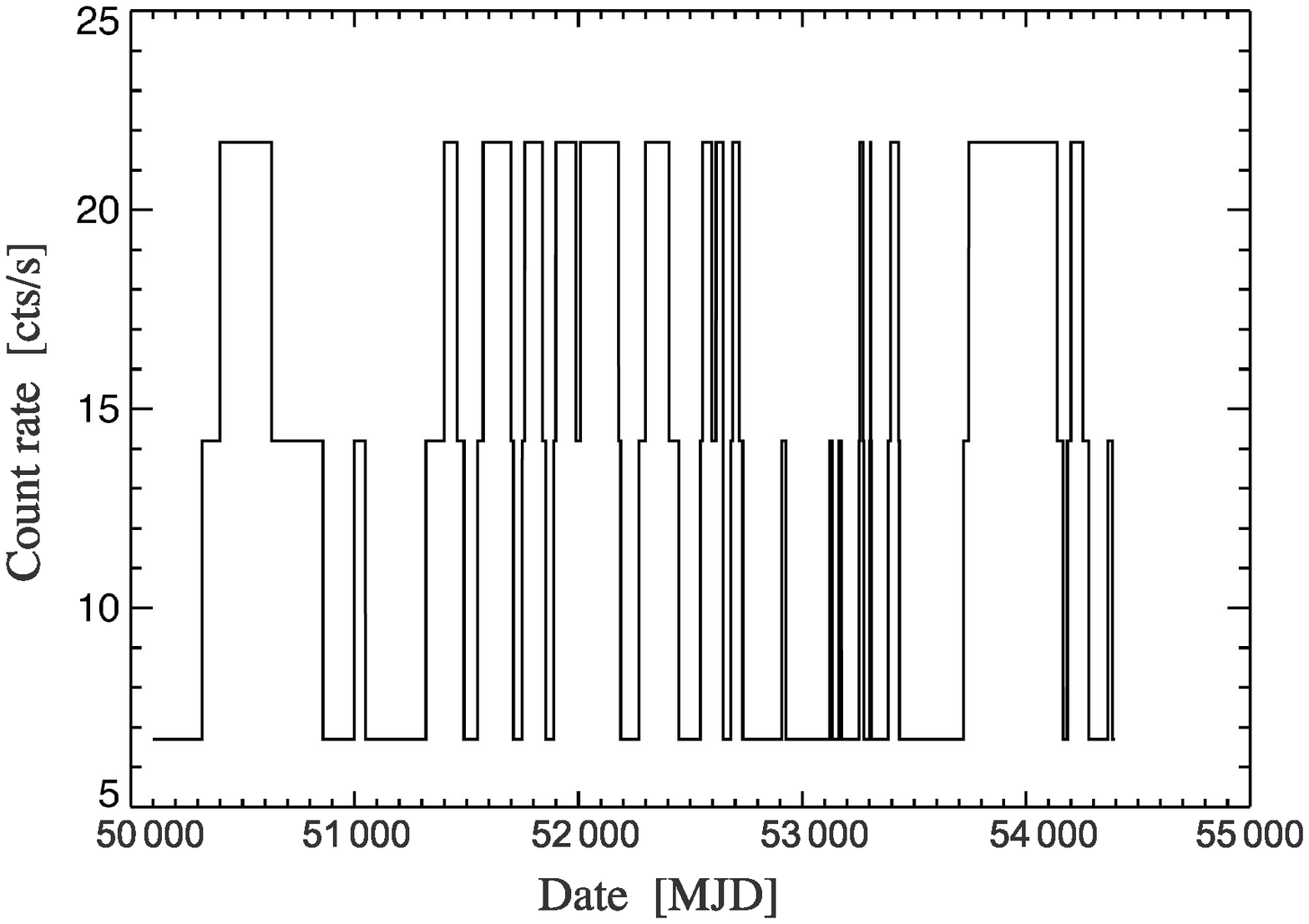}
\includegraphics[width=8cm]{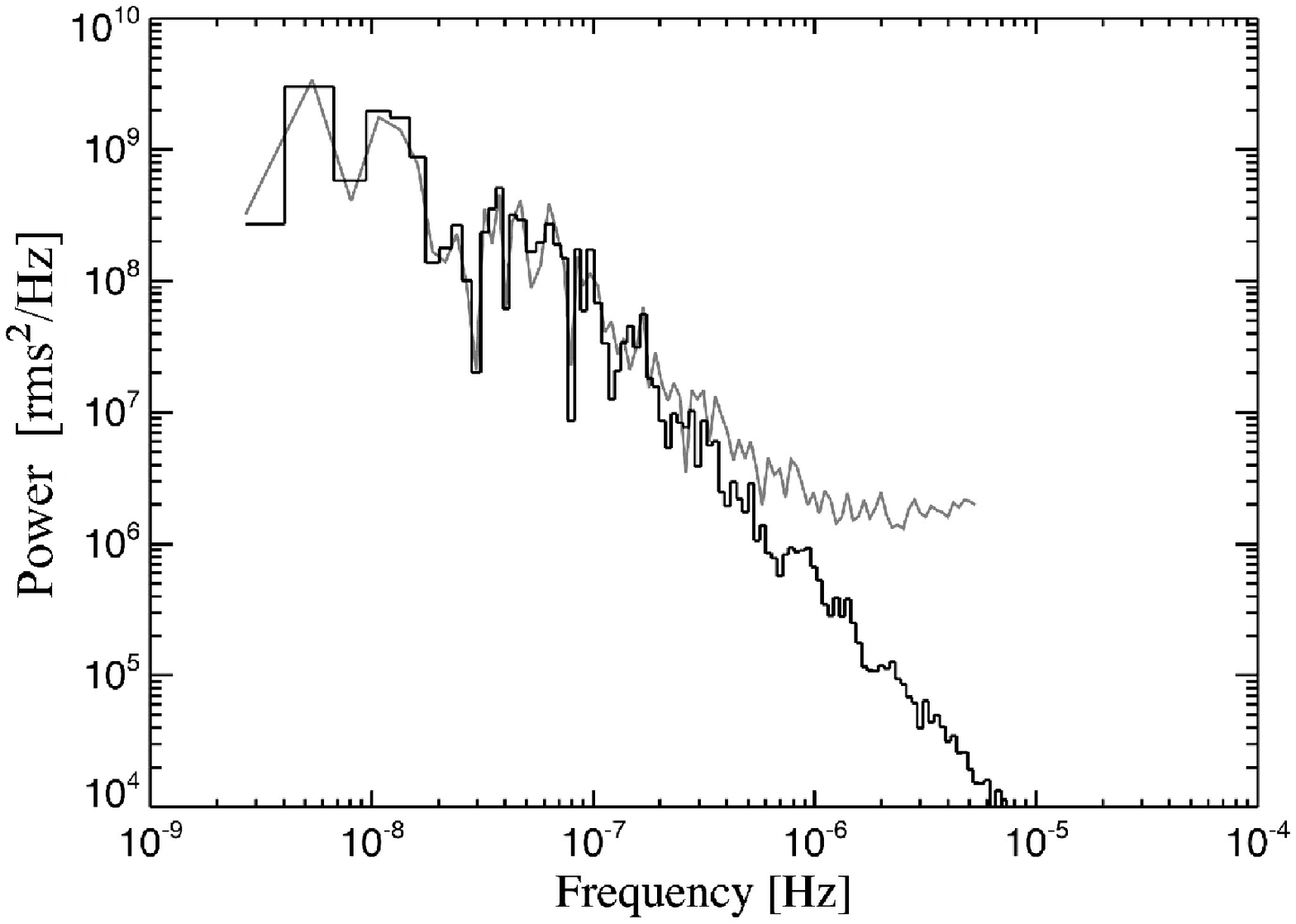}
\caption{Results of testing the contribution of state transitions to the PDS of {\cx}. {\it Left panel:} The artificial lightcurve created with only three levels, corresponding to low, intermediate and high states. {\it Right panel:} The PDS of the artificial data (black curve) overlaid on that of the real lightcurve (grey curve). The two curves agree almost completely below $\sim 10^{-7}$ Hz.}
\label{statepds}
\end{figure*}

We now calculate the PDS using the method described in Sect.~\ref{uneven}. To check for differences between the states, we first split the data into segments based on the {\asm} count level: hard state (count rate $\sim10$ cts/s) and soft state ($\sim20$ cts/s). A PDS was calculated for each segment, and all PDS of the same state were then averaged for the final result. 
As for the higher frequencies, the PDS in both states is well described by a power-law. The index in the hard state is $-0.93\pm0.09$ and for the soft state $-0.78\pm0.07$. However, these PDS are distorted by the effects of the data sampling, and the indices derived do not give a true representation of the intrinsic variability. From our approach using simulations, we find that the power-law of the intrinsic PDS is steeper: the observed results correspond to an intrinsic power-law index of $-1.3$ in the hard state and $-1.5$ in the soft state. 
The relative error of the corrected indices are roughly the same as for the uncorrected ones, i.e., about 10\%. The difference in index between the two states is therefore not significant.

\subsubsection{State transitions}

We have also combined all data, and calculated the PDS for the entire ASM lightcurve. The resulting index is in this case steeper than when either of the two states is treated individually. 
Looking at the {\asm} lightcurve, it is clear that the state transitions are responsible for much variability on long time-scales.  
To determine the contribution of the transitions to the PDS we produced an artificial lightcurve with only three levels, corresponding to the hard, intermediate and soft states. This removes all variability \emph{within} each state, isolating the power produced by the transitions themselves. The resulting lightcurve is shown in Fig.~\ref{statepds} (left panel).
The artificial time series was evenly sampled, with a bin length of 0.1 days and with no data gaps. We then computed the PDS for our artificial data and compared to the real PDS. The two curves are shown overlaid in the right panel of Fig.~\ref{statepds}.

In the figure we see that there is almost complete agreement between the real and artificial data at lower frequencies. We stress that we have not performed any fitting of the two curves. Nevertheless, they follow each other almost completely up to $\sim 10^{-7}$ Hz. The PDS of {\cx} below this frequency can therefore be explained simply by the variability of the ASM flux connected with the state transitions. It is also clear that the time sampling of the ASM data is not responsible for the structure of the PDS at low frequencies since these effects are not present in our artificial state lightcurve.

Using the ASM data we have looked for possible correlations between the time that Cyg X-3 stays in a soft state and the length of the preceding and of the following hard and intermediate state. With the 15 soft states available for this analysis no obvious deviation from a random scatter was found.

We also analysed the distribution of the length of time that the source remained in a hard or soft state. The mean lengths of the 19 hard and 15 soft states were 103 and 98 days, respectively. The Cumulative distribution function for each of the states was compared with that of the waiting times for a Poisson process with the same mean value. The Kolmogorov-Smirnov D-value was 0.125 and 0.131 with corresponsing significances of 0.907 and 0.943, respectively. Both distributions are therefore consistent with a process where the probability of a transition at any time is independent of the time since the last transition. The fact that the distributions are well explained by a Poisson process does not exclude other possibilities.

\subsection{The complete picture}

To complement the data from the {\asm} and {\pca} instruments, we use the data from {\exo} for the intermediate time-scales ($10^{-5}-10^{-2}$\,Hz). As expected, the PDS at intermediate frequencies is well described by a broken power-law: at lower frequencies the index matches that of the ASM data, and at higher frequencies the power-law steepens, matching the PCA data.

Of the 19 observations used, 3 are in the hard state, 1 is in the intermediate state and 15 are in the different soft states (which we again combine together into one PDS). The resulting indices are presented in Table~\ref{exotable}. We find a slightly higher index in the hard state than in the soft and intermediate states. Although the difference is not significant, this matches the trend seen in the PCA data, where the hard state shows a slightly less steep power-law.

\begin{figure}
\includegraphics[width=8cm]{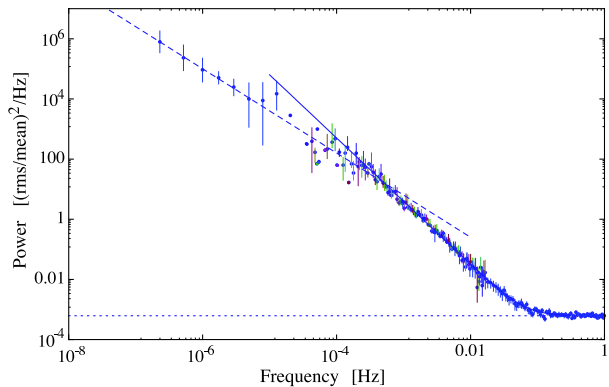}
\caption{Broadband power spectrum of Cyg~X-3. The solid line is a power-law with index $-2.1$, fit to the {\pca} data, and the dashed line is a power-law with index $-1.3$ matching the {\asm} data. The dotted line indicates the Poisson level.}
\label{broadband}
\end{figure}

In all our {\it EXOSAT} observations we find signs of a break around $10^{-4}$\,Hz. The break is seen
in the PDS of both the uncorrected lightcurves and of the lightcurves with orbital demodulation applied. The limited frequency range below the break in the {\exo} data together with the orbital modulation and demodulation make it difficult to determine the low-frequency index. However, the long (110\,ks) intermediate state observation presented in \citet{wil85} does allow a fit at lower frequencies. Analysing this observation alone, we find an index of $-1.08\pm0.53$ below $10^{-4}$\,Hz. From our simulations we can determine that the break itself not significant; however, extrapolating the PDS at low and high frequencies gives an intersection around $10^{-4}$\,Hz.
Fig.~\ref{broadband} shows the total combined broad-band PDS covering the frequency range $10^{-7}-0.1$\,Hz. The {\asm} and {\exo} data combine all soft states, and the {\pca} data shown is from the soft non-thermal state.

\begin{table}
\begin{center}
\begin{tabular}{l l}
\hline \hline
State & \hspace{5mm}Index \\
\hline
Hard & $-1.78\pm0.07$ \\
Intermediate & $-1.9\pm0.1$ \\
Soft & $-2.00 \pm 0.04$ \\

\hline
\end{tabular}
\caption{Fitted power-law indices above $10^{-4}$\,Hz in three states using {\exo} data.}
\label{exotable}
\end{center}
\end{table}

\section{Discussion}

\subsection{The PDS as a state indicator}

In many XRB sources, the power spectrum at high frequencies changes with the spectral state of the source. As shown by several authors \citep[e.g.,][]{bel02,axe06,don07} studies of the PDS complement spectral analysis, and may in some cases provide a clearer indication of state than the spectral energy distribution. Unlike other X-ray binary sources, for exmple Cyg~X-1, the PDS of {\cx} do not show large variations between different spectral states. The PDS in the different soft states are indistinguishable from each other and the only possible difference we may see is a slight tendency for a less steep power-law in the hard state. 

The case for a true state transition in {\cx} would be considerably strengthened had quasi-periodic oscillations (QPOs) been detected in the hard state PDS, as in other sources, but we see no evidence of such features \citep[nor of any transient oscillations, as in][]{vdk85}. In the high-mass black hole binary Cyg~X-1, these features dominate the hard state PDS above $\sim0.01$\,Hz, with the lowest frequency feature peaking at 0.1--1 Hz. In this range the PDS of {\cx} is however below the Poisson level which would prevent such features from being detected. Although the origin of the QPO features is not yet understood, many models predict an inverse scaling with mass \citep[i.e., a scaling with the Keplerian frequency, see for example][]{vdk04}. 

A generally accepted model for the accretion flow around the compact object is a two-phase flow, with an outer cool thin disc that is truncated at some radius and replaced by a hot inner flow. The truncation radius is determined by the mass accretion rate -- at high accretion rates the thin disc may extend in to the innermost stable orbit \citep[for a thorough review of the model, see][]{don07}. This model provides a natural candidate for a specific radius where the frequencies are realised: the truncation radius. The low-frequency feature can be estimated using the viscous time-scale (the longest time-scale) at the truncation radius, giving \citep{don07}:
\begin{equation}
\nu\sim0.2\left(\frac{r}{6}\right) ^{-1.5}\left(\frac{m}{10}\right)^{-1}\;{\rm Hz}\,,
\end{equation}
where $r$ is the truncation radius in units of gravitational radii and $m$ the black hole mass in solar masses. \citet{hja08b} concluded that for the transition to a hard state in {\cx} to occur at the same Eddington luminosity as in other non-transient sources, the mass of the compact object must be $\sim30\;\msun$, three times that of Cyg~X-1. Scaling up the low frequency component in the PDS of Cyg~X-1 for this mass , we would expect this component to appear at $\sim 0.03-0.3$ Hz. In this range, it {\it could} be visible if it is there. However, the exact truncation radius and blackbody temperature in {\cx} are poorly known due to both absorption and emission features in the wind that affects the spectrum strongly at low energies \citep[e.g.][]{sz08}. According to the best-fit models of \citet{hja08b} the disc in the hard state of {\cx} is truncated further in than in the hard state of Cyg~X-1. It is therefore likely that the component would show up in the higher end of the frequency range,  which is again close to where the variability drops below the Poisson level. In conclusion, the non-detection of any QPO features in the hard state PDS of {\cx} neither confirms nor argues against a true state transition in the source.

\subsection{Effects of the wind on the PDS}
\label{wind}
Continuing the comparison with other XRBs, it is clear that the power-law in {\cx} is quite steep - the most common value in other sources is an index close to $-1$ \citep[at least in the soft states where the power-law dominates; see][and references therein]{vdk04}. An interesting question is therefore what process could give rise to this PDS. As the system is enshrouded in heavy stellar wind, it is tempting to explain the PDS as the result of scattering in this medium. 
Such a scenario would predict that frequencies above the random walk time-scale are suppressed in the PDS, much like the higher frequencies in {\cx}. The explanation was suggested by \citet{ber94} to explain the weak or missing high-frequency variability in {\cx}. If the wind optical depth does not  vary much between the states of the source, this could also explain the more or less constant appearance of the PDS.

This scenario requires the Wolf-Rayet wind to be optically thick to scattering. The typical scattering time-scale for a spherical homogeneous scattering cloud of radius $R$ and optical depth $\tau$ is $0.5 R \tau / c$ \citep{kyl89}. The  $\sim-2$ index seen in the PDS of the PCA data continues to $10^{-4}$\,Hz. Associating the break at $10^{-4}$\,Hz with the scattering time-scale and assuming a radius of  the dense wind region of $10^{12}$\,cm (an estimate based on IR observations; \citealt{fen99}), would however require an unrealistic optical depth of $\sim 600$. 

\citet{sz08} modelled the Wolf-Rayet wind including absorption and emission features. They find an average $\tau\approx 0.3-0.4$. This value gives a scattering time-scale of the order of only a few seconds for a radius of $10^{12}$\,cm, or requires a radius of $3\times10^4$\,R$_{\sun}$ to explain the break in the PDS. In \citet{hja08b} it was suggested that the low electron temperature in the hard state of {\cx} may be a result of Compton down-scattering in an optically thick wind, either from the Wolf-Rayet star or in the form of a disc wind/outflow (this effect was however not included in the modelling of the X-ray spectrum). Such a wind would have to have an optical depth of $\tau\sim$3--5 to down-scatter typical 100 keV photons to the observed 20 keV peak. However even such a dense wind would still have to be extremely large, $\sim2500$\,R$_{\sun}$. 

Thus, while scattering in a nearby surrounding medium may explain the lack of high-frequency variability ($\ga0.1$\, Hz) as suggested by \citet{ber94}, the $\sim-2$ index of the PDS cannot be attributed to random walk in the Wolf-Rayet wind or any local scattering medium. 

Allowing for a very large size of the scattering cloud does not solve the problem. Although a large scattering halo has been detected around {\cx} and the time delay from scattered X-rays used to determine the distance to the source \citep{pre00}, this is the result of scattering in interstellar dust. The optical depth in this medium is thus $\ll1$, and the halo is not a sign of the source being extended.  However, an extended feature has been found around the {\cx} system \citep{hei03}. It does not surround the entire source, and has been interpreted as a localized remnant of earlier outflows. Recently, the feature has been shown to display flux and phase correlations with {\cx}, making such an interpretation less likely \citep{mcc08}. A possibility is that the feature is the result of a small scattering cloud within 2 kpc of {\cx}, and can therefore not explain the high-frequency power spectrum.

Our results thus rule out the steep power-law as being due to scattering effects in a Wolf-Rayet wind, disc wind, local or line of sight matter, with any reasonable physical parameters.

\subsection{Long time-scales and transitions}

The PDS of many XRBs are well described by a power-law at low frequencies. In a survey of low-mass XRBs, \citet{gil05} found that most of the studied systems displayed a power spectrum with index close to $-1.3$ at low frequencies. The physical origin of this variability is however not yet determined. A commonly adopted framework is that of \citet{lyu97}, where disturbances in the outer region of the disc propagate
inwards. Faster variations thus get superposed on slower variations, and the emergent power spectrum will have the shape of a power-law. A physical model based on this scenario was proposed by \citet{kin04}, and \citet{tit07} showed that a two-phase model (cool disc and hot flow) could successfully fit the broadband PDS of Cyg~X-1 and Cyg~X-2. The models predict a break in the frequency spectrum at time-scales corresponding to the longest variations, and a flat (white noise) PDS below this. 

\citet{gil05} associate the longest time-scales in the accretion disc with the viscous time-scale at its
outer boundary. We do not detect any break or flattening at the lowest end of our frequency range, that could be associated with an outer disc radius. On long time-scales, it is clear that the variability introduced by the state transitions dominates the power spectrum, making such a feature hard to detect in the PDS of the complete ASM lightcurve (cf. Fig~\ref{statepds}) if the disc is larger than half the orbital separation. Since {\cx} is a wind-accreting system, it is likely that the disc significantly smaller. A feature in the PDS connected to the outer edge of the disc could still be present at higher frequencies. This possibility will be discussed in the next section. 

Even if dominant at our lowest frequencies, it is clear that the power from the transitions alone does not explain the PDS at higher frequencies, showing that we are seeing intrinsic variability from the X-ray emitting region. The PDS above $10^{-7}$\,Hz can be explained within the framework presented above, with disturbances in the accretion disc propagating inwards and ultimately modulating the X-ray flux.
If the steep power law at high frequencies is not due to any scattering effect, it means that the PDS is intrinsically steep. In the model of \citet{lyu97} a power-law index of $\sim-1$ arises if perturbations in the disc are similar at all radii. If the perturbations are greater at large radii, the resulting power spectral index is steeper. This can be coupled to the disc being thickened at large radii, as proposed for the suggested accretion geometries in \citet{hja08b}.

Our results further show that the probability of a state transition in {\cx} is not dependent on the amount of time since the previous transition. This is contrary to what one might expect for transient systems, where an outburst consumes a large fraction of the accretion disc. The time-scale of an outburst is therefore dependent on the amount of matter in the disc, which should correlate with the time spent in quiescence, when the disc is built up. The state transitions in {\cx} thereby seem more directly coupled to the mass loss rate of the companion star. This means that the viscous time-scale for a reconfiguration of the accretion flow as a response to a change in the accretion rate is shorter than the typical time-scale for variations in the mass-loss rate of the companion (or capture rate of the compact object).

\subsection{The break in the power spectrum}

Although the break around $10^{-4}$\, Hz seen in the {\exo} PDS is not formally significant, the fact that it is consistent with the intersection of the power-laws from the (high-frequency) PCA and (low-frequency) ASM power spectra speaks in favour of the break being real. The same feature was seen in the results of \citet{wil85}, who analysed a single long {\exo} observation. The orbital modulation has been compensated for in both our analysis and that of \citet{wil85}. Nevertheless, the break is quite close to the time-scale of the orbital period which is slightly suspicious. We note that the demodulation in all cases had a negligible influence on the PDS below $10^{-5}$\,Hz, so the main uncertainty is not the existence of a break, but its precise location.

If the break is interpreted as the feature connected with the viscous time-scale at the outer edge of the disc, this would mean a very small accretion disc. The corresponding radius is approximately $3\times10^{-2}\;{\rm R_{\odot}}$, which is unreasonably small, and this interpretation can therefore be ruled out.

It is interesting to further compare our results to the analysis of low-mass XRBs presented by \citet{gil05}. 
They found that all compact systems in their study showed a feature in the PDS close to the orbital frequency. These systems also have low mass ratios, \mbox{$q=M_2/M_1\la0.3$}. 
\citet{gil05} argued that for low mass ratios, it is possible to excite a 3:1 resonance in the disc. The resonance occurs if the angular frequency of the orbital motion of a particle in the disc is commensurate with the angular frequency of the orbital motion of the secondary and will lead to asymmetry of the disc and a possible feature in the PDS. For systems with a higher mass ratio, the disc is tidally truncated before reaching the resonance radius.

Tying the break at $10^{-4}$\,Hz observed in {\cx} with the 3:1 resonance indicates that the mass ratio in this system is low, $q\la0.3$. Although the mass of the companion Wolf-Rayet star is poorly known, it is expected to be $\ga 10\,\msun$ \citep[e.g.,][]{cro07}, giving a black hole mass of $\sim30\,\msun$. This value is in agreement with that suggested by \citet{hja08b}.

\section{Conclusions}

We have conducted a systematic analysis of the PDS of {\cx} over the frequency range $10^{-9}$--1 Hz. For the first time, we compare the PDS in all states of the source.

On short time-scales, we confirm previous results that the PDS is well described by a power-law with index close to $-2$. By comparing the variations within each state to that between states, we find that only the hard state index shows a tendency to deviate from this value. We do not find any evidence of QPO features in the PDS, but this neither confirms nor contradicts the presence of a truncated disc in the hard state. Our conclusion is that the PDS is not a good indicator of state in {\cx}.

On longer time-scales we find that the variability is dominated by the state transitions: the PDS below $\sim 10^{-7}$ Hz can be completely explained by this pattern. We also find that there is no correlation between the probability for a state transition and the length of time spent in the state, suggesting a direct feedback between mass-capture rate and spectral state.

 Combining datasets, we were able to create a power spectrum covering the entire $10^{-7}$--\,0.1\,Hz range, and found evidence for a break in the PDS around $10^{-4}$\,Hz. This may be interpreted as evidence for the existence of a tidal resonance in the accretion disc around the compact object.

\section*{Acknowledgments}

This research has made use of data obtained through the High Energy Astrophysics Science Archive Research Center (HEASARC) Online Service, provided by NASA/Goddard Space Flight Center. LH acknowledges support from Anna-Greta and Holger Crafoord's Fund, through The Royal Swedish Academy of Sciences.

\end{document}